\documentstyle[preprint,aps,prl,psfig]{revtex}
\begin{document}
\topmargin -0.05cm
\title{Origins of Chevron Rollovers in Non-Two-State Protein Folding Kinetics}
\author{H\"useyin K{\footnotesize{AYA}}  and Hue Sun C{\footnotesize{HAN}}}
\address{Protein Engineering Network of Centres of Excellence,\\
 Department of Biochemistry, and \\
 Department of Medical Genetics \& Microbiology \\
 Faculty of Medicine, University of Toronto \\
 Toronto, Ontario M5S 1A8, Canada}

\maketitle

\begin{abstract}

Chevron rollovers of some proteins imply that their 
logarithmic folding rates are nonlinear in native stability.  
This is predicted by lattice and continuum G\=o models to 
arise from diminished accessibilities of the ground state     
from transiently populated compact conformations under strongly
native conditions. Despite these models' native-centric         
interactions, the slowdown is due partly to kinetic trapping  
caused by some of the folding intermediates' nonnative 
topologies. Notably, simple two-state folding kinetics of     
small single-domain proteins are not reproduced by common 
G\=o-like schemes.

\end{abstract}
\eject

The physical basis of protein folding is a central unresolved puzzle 
in molecular biology. Recently, much advance in protein folding has
originated from experiments on small single-domain proteins [1] with 
simple two-state folding and unfolding kinetics typified by that of 
CI2 [2], with features including: (i) single-exponential relaxation, 
(ii) the logarithmic folding and unfolding rates ($\ln k_{\rm f}$ 
and $\ln k_{\rm u}$) at constant temperature being essentially linear 
in chemical denaturant (urea or GuHCl) concentration, i.e., both arms 
of the ``chevron plot'' [3] are linear, and that (iii) the equilibrium 
ratio of native to denatured conformational population $K\equiv$ 
[native]/[denatured] $=k_{\rm f}/k_{\rm u}$. What form of intrachain 
interactions might give rise to such remarkable behavior is a question 
of fundamental biophysical interest.

Other proteins' corresponding properties are more complex. Often this is 
manifested by significant deviations [4--9] from the above linearities,
i.e., they exhibit chevron rollovers [10]. We refer to their kinetics
as non-two-state. Examples of such behavior include barnase [4,8], 
ribonuclease A [5], hen lysozyme [6], and U1A [7]. 
The present operational definition of non-two-state kinetics 
encompasses what some authors called ``two-state'' (though not ``simple'') 
when conditions (i) and (iii) above are satisfied but not (ii) [7,8].
Chevron rollover can also be brought about by mutation, as in S6 [7] and 
BPTI [9]. Thus, rather than an aberration, chevron rollover is quite 
ubiquitous. Therefore, ascertaining its physical origins should 
provide important clues to protein energetics. 

Chevron rollovers have been attributed to peculiarities of 
intermediates or transition states on postulated free energy profiles 
[4--8,10], or front factors' sensitivity to folding conditions [11]. 
Yet these phenomenological considerations do not pinpoint 
the physical processes involved. In this Letter, physical mechanisms 
underlying chevron rollovers are addressed directly by examining a 
multitude of trajectories from several protein chain models.

The recent discovery of a remarkable correlation between contact order 
and folding rates of simple two-state proteins [12] has led to extensive 
studies of G\=o-like protein models [13--17]. Hence,
a natural question is whether common G\=o-like constructs
do predict simple two-state kinetics. Somewhat surprisingly, our investigation 
thus far indicates that this may not be the case. Instead, chevron rollover 
emerges as a conspicuous feature in both lattice [11] and continuum [17]
G\=o models. This suggests that common native-centric [16] chain 
constructs can be useful for elucidating the polymer mechanisms of chevron 
rollovers, even though they may not be entirely adequate for simple 
two-state proteins. Pursuing this logic, we now analyze a thermodynamically 
cooperative [16] 48mer three-dimensional lattice G\=o model [14]. This model 
had notable impact on recent appraisals [18] of the energy 
landscape views of protein folding [19], but its chevron behavior 
has not been investigated.

Each native contact in this model has a favorable energy 
$\epsilon$ ($<0$),
nonnative contacts have zero energy. Folding/unfolding kinetics are modeled 
by Metropolis Monte Carlo (MC) dynamics with the same set of 
elementary chain moves as in [14]: End moves are attempted for the two 
chain-end monomers.  Corner and crankshaft moves are attempted for
other monomers with 70\% and 30\% probability respectively. Time
is measured by the number of attempted MC moves; $Q$ is fractional number 
of native contacts [15,17].

The chevron plots for this G\=o model and a closely related model are 
shown in Fig.~1. When relaxation is single-exponential (see below), 
$k_{\rm f}$ or $k_{\rm u}$ $=$ 1/MFPT. Most of the MFPTs 
here are averaged from at least 1,000 trajectories, except for a narrow
$\epsilon/k_{\rm B}T$ range 
around the transition midpoint where kinetics are relatively
slow (100--200 trajectories 
each) and for folding initiated from Fig.~2(d) (200 
trajectories each). In these models, the free energy $\Delta G_{\rm u}$ 
of unfolding 
to the open conformations ($Q\le 6/57$) is essentially linear in 
$\epsilon/k_{\rm B}T$. Thus we model denaturant concentration changes
by varying $\epsilon/k_{\rm B}T$ [11,20]. Adding repulsive nonnative contact
energies to a G\=o model [20,21] does not appear to have a significant impact 
on the chevron behavior. Fig.~1 shows dramatic chevron rollovers of the folding 
arms and very slight rollovers of the unfolding arms for both models,
with maximum folding rates at $\Delta G_{\rm u}/k_{\rm B}T$= 14.2 (G\=o) and 
16.2 (G\=o plus repulsion). Fig.~1 indicates that deviations from simple 
two-state behavior can be difficult to discern under weakly native 
conditions [22]. To facilitate comparison with experiments, we 
characterize folding rollover by the difference between the hypothetical 
simple two-state $\ln k_{\rm f}^{\rm 2-s}$ (inclined dotted lines in 
 Fig.~1) and the actual (simulated) folding rate $\ln k_{\rm f}$ $\approx$ 
$-\ln({\rm MFPT})$ at three representative values of native stability
$\Delta G_{\rm u}$,
spanning a range typically covered by real proteins. Here, for 
$\Delta G_{\rm u}/k_{\rm B}T=$ (5, 10, 15), the logarithmic rollover ratio 
$\ln (k_{\rm f}^{\rm 2-s}/k_{\rm f})$ $=$ ($0.16$, $1.08$, $2.96$) 
for the G\=o model and ($0.32$, $1.36$, $3.12$) for the model with 
repulsive interactions. These ratios are not dissimilar to the corresponding 
$\ln (k_{\rm f}^{\rm 2-s}/k_{\rm f})$ $\approx$ ($0.77$, $2.42$, $4.28$) 
for wildtype barnase at 25$^\circ$C and pH 6.3 [4]. Under these
conditions, a maximum folding rate 
was not observed for barnase [4]. However, if a 
quadratic dependence [7,23] of $\ln k_{\rm f}$ vs. denaturant is assumed
for barnase (c.f. [8]), a maximum folding rate $\approx 230$ s$^{-1}$ may be 
extrapolated to occur at an hypothetical 
$\Delta G_{\rm u}\approx 40 k_{\rm B}T$ which is
much more stable than the $\Delta G_{\rm u}\approx 18.0 k_{\rm B}T$ 
at zero denaturant.

\vskip -.01cm

 Fig.~2(a) provides the G\=o model's conformational distributions 
at different native stabilities.
Under mildly native conditions 
($\Delta G_{\rm u}<10 k_{\rm B}T$), the free energy profiles have
a barrier between the native and denatured minima. 
Under more strongly native conditions 
($\Delta G_{\rm u}> 15 k_{\rm B}T$), their shapes are suggestive of
downhill folding [24]. The analysis of first passage time distribution 
[16,25] in Fig.~2(b) indicates that folding kinetics is approximately 
single exponential [$\ln P(t)$ linear in $t$] under mildly native
conditions, consistent with the 
observed single-exponential folding kinetics for ribonuclease A 
when double-jump experiments were used to eliminate the effect of 
{\it cis/trans} proline isomerization [5]. 
The behavior of wildtype barnase is
similar: Folding is fast and single-exponential for
the majority of the chains ($\approx 80\%$), the rest 
belongs to a slow-folding tail caused by proline isomerization [4].
However, when modeling conditions are strongly native (corresponding
conditions may not always be experimentally achievable [11]), 
folding kinetics is not single-exponential [circles and 
squares in Fig.~2(b)]. The onset of this behavior
occurs approximately when the $-\ln [P(Q)]$ profile becomes downhill
and where folding rate is maximum (c.f. Figs.~1 and 2) [11]. It would 
be instructive to ascertain whether 
this specific model prediction applies to real proteins.

A closer examination of the model folding trajectories indicates
that the slowdown leading to folding-arm chevron rolloves
arises from transiently populated compact non-ground-state conformations 
because these
folding ``intermediates'' have lifetimes that increase with increasingly
native conditions (Figs.~1, 3 and 4). Examples are shown in Fig.~3(b--d). 
Once one of these structures is adopted under 
strongly native conditions ($\epsilon/k_{\rm B}T>1.55$), it takes 
longer on average to reach the ground state if intrachain native 
contacts are more favorable (Fig.~1). Folding trajectories 
under strongly native conditions are qualitatively different
from that under milder conditions [Fig.~4(a--c)]. At the transition
midpoint [Fig.~4(c)], interconversions between $Q\approx 0.2$ and
$Q\approx 1.0$ are sudden and sharp. Relatively little time is spent at
intermediate $Q$ values. As $\epsilon/k_{\rm B}T$ increases, however, certain 
conformations with intermediate $Q\approx$ $0.6$--$0.8$ are 
frequented more.  
Even under mildly native conditions, their impeding 
effects on folding kinetics is already apparent from the event in Fig.~4(b)
depicting the chain bounces back to $Q\approx 0.2$ after achieving 
$Q\approx 0.8$. 
But the lifetimes of these ``intermediates'' are brief compared 
to that of the open unfolded conformations. Hence folding remains 
approximately single-exponential [triangles in Fig.~2(b)]. 
However, when conditions become more strongly native [Fig.~4(a)], 
some folding trajectories are dominated by ``intermediates,'' leading 
to a significant reduction in average 
folding time. But even under these circumstances it is
still possible to fold quickly [Fig.~4(a), inset]. 
Consistent with Fig.~2b (circles), this separation of time scales means 
that folding is no longer single-exponential.

In contrast to a previous report that no ``entangled misfolded state''
was observed during the folding of this particular G\=o model [14], 
 Fig.~3(b) exhibits an overall 
nonnative topology. For Fig.~3(c), the left side of the conformation 
is native, but the right side is substantially nonnative. Hence these 
conformations are kinetic traps in that they cannot reach the ground state 
without first open up somewhat by breaking some existing 
favorable contacts. Notwithstanding possible artifacts of lattice models 
 [Fig.~3(d)], this basic physical requirement rationalizes folding-arm
chevron rollover because favorable contacts contributing to the 
meta-stability of these traps are increasing difficult to break with 
stronger $-\epsilon/k_{\rm B}T$. 

This prediction appears to be robust over a range of lattice and 
continuum coarse-grained models [Fig.~4(d--e)] that exhibit chevron rollovers. 
 Fig.~4(d) shows that folding 
of a recent lattice model with residue-based as well as 
native-centric interactions are similar to that in Fig.~4(b).
 Fig.~4(e) shows that 
folding of a continuum C$_\alpha$ model 
under mildly and strongly [inset in (e)] native conditions are 
very much similar, respectively, to the corresponding
lattice results in Fig.~4(b) and 
(a). The trajectories in Fig.~4(f) from a
continuum model [17] with desolvation barriers [26] 
also show that
intermediate $Q$ values are more prominent during the folding process
when conditions are more strongly native [inset in (f)].

A maximum (or ``optimal'') folding rate similar to those in Fig.~1 have
been observed in many models (e.g., [19,20,21,23,25]) since it
was first noted in HP model simulations more than a decade ago [27]. This 
feature arises from a competition between a stronger driving force for 
folding and the onset of glassy dynamics under strongly native conditions
(see note added in proof of [20]). However, until recently [11,17],
the connection between this theoretical phenomenon and chevron rollover 
has not been recognized. Perhaps this is because the maximum folding rate 
often occurs near the transition midpoint for less cooperative models, 
and hence its relationship with chevron plots is less obvious.
In contrast, the models studied here possess proteinlike thermodynamic 
cooperativity [16]. Several basic principles now emerge: 
(i) Rollovers can arise from kinetic trapping [6,9,20]; but folding 
relaxation remains approximately single exponential
when trapping effects are mild [Fig.~2(b)]. (ii) We have rationalized
rollovers phenomenologically by front factors that depend on
$\Delta G_{\rm u}$ [11,17]. Physically, this dependence is 
likely caused by trapping and unfolding (barrier recrossing) from 
transiently populated compact non-ground-state conformations (Fig.~4).
These predictions are testable by experiments.
(iii) The chevron rollovers in the G\=o models presented suggest 
strongly that, contrary to expectation, G\=o-like pairwise additive 
interactions are insufficient [17] to capture the remarkable kinetics of 
small single-domain proteins [1]; further research is necessary to 
ascertain the physical origin of their simple two-state 
cooperativity.

The authors thank Walid Houry for a very helpful discussion. 
This work was supported in part by Canadian Institute of Health 
Research grant MOP-15323.


\par\vfill\eject




\par\vfill\eject


\noindent
{\bf Fig.~1} $\quad$ MFPT is mean first passage
time, $k_{\rm B}T$ is Boltzmann constant times absolute temperature. 
Circles are for the 48mer G\=o model [14]. Squares are data from a model 
with the same ground-state conformation and attractive energy $\epsilon$ 
for each native contact but with an additional repulsive energy $-\epsilon$ 
for each nonnative contact (as in the HP+ model [20] and Ref.~[21]). 
Folding (open symbols) starts from a random self-avoiding walk, first 
passage is achieved when $Q=1$. Unfolding (filled symbols) starts from 
the $Q=1$ ground state, first passage is achieved when $Q\le 6/57$
because the free energy 
minimum of the denatured states is at $Q\approx 6/57$. Solid dashed 
curves are mere guide for the eye. The vertical dotted lines mark the 
two model's thermodynamic transition midpoints. The two pairs of V-shape 
dotted lines are hypothetical simple two-state chevron plots [11] based upon 
$\Delta G_{\rm u}$ between $Q=1$ and $Q\le 6/57$ obtained by
histogram techniques from sampling around the transition midpoint.
The triangles, asterisks, and diamonds are $-\ln({\rm MFPT})$ values for 
G\=o-model folding initiated (at $t=0$) respectively from the compact 
conformations (b), (c), and (d) in Fig.~3. Arrows indicate the 
$\epsilon/k_{\rm B}T$ values considered in Fig.~2. 
\\

\noindent
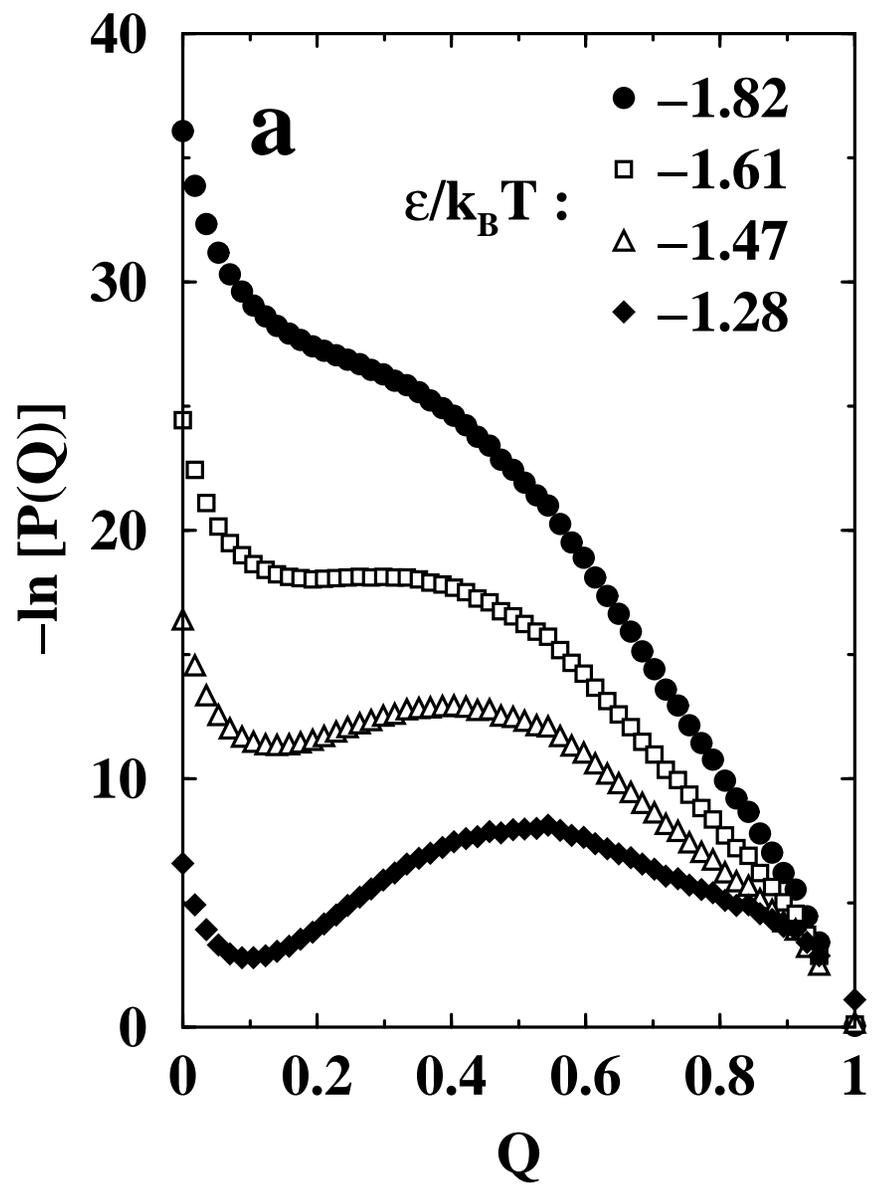
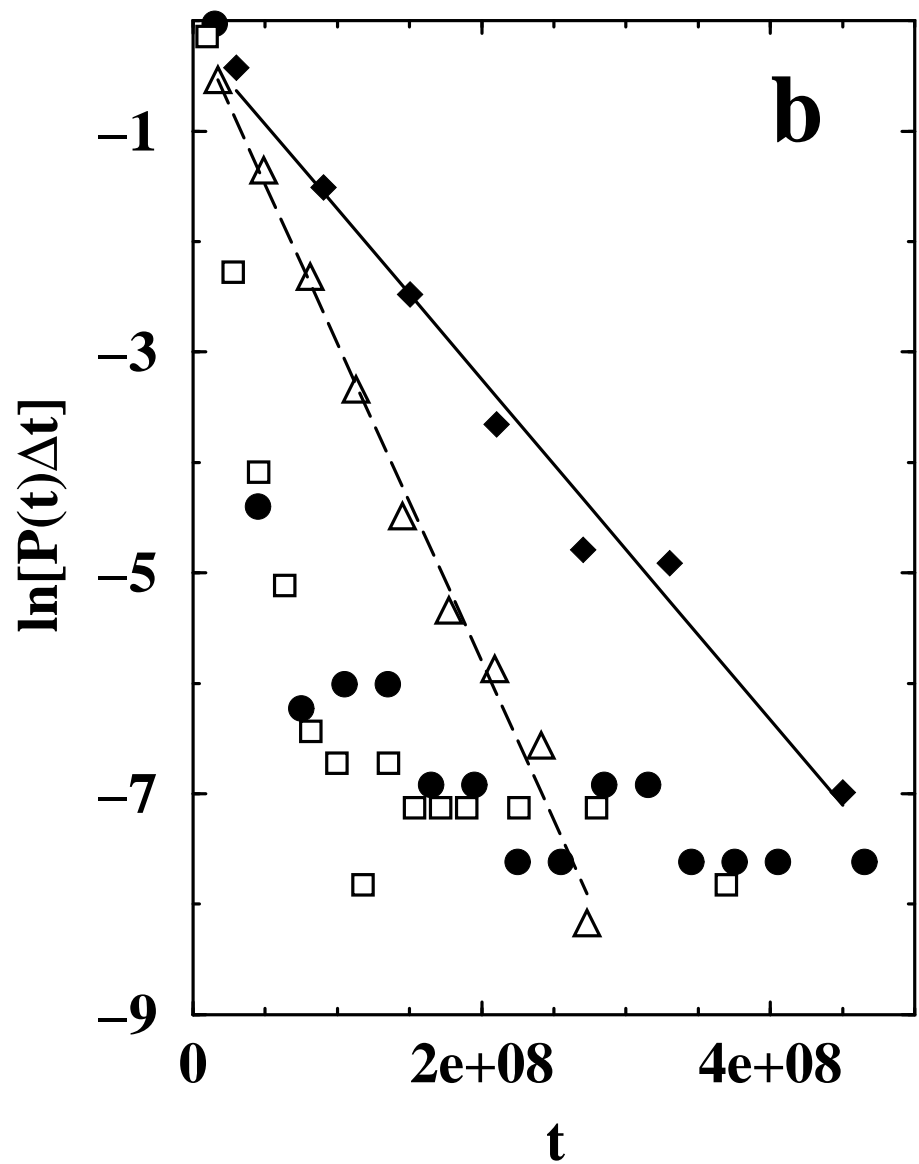
{\bf Fig.~2} $\quad$ (a) Free energy profiles for the G\=o model at the 
$\epsilon/k_{\rm B}T$ values indicated (c.f. Fig.~1). $P(Q)$ is Boltzmann 
population distribution over $Q$. [Note that $P(Q)=0$ for $Q=$ $55/57$ and 
$56/57$.] (b) $P(t)\Delta t$ is the fraction of folding trajectories 
with $t-\Delta t/2<$ first passage time $\le t+\Delta t/2$ [17], plotted 
in different horizontal scales for different $\epsilon/k_{\rm B}T$s 
[symbols as in (a)] to enhance clarity. For $\epsilon/k_{\rm B}T=$ 
$-1.82$, $-1.61$, $-1.47$, and $-1.28$ respectively, 2,500, 2,030, 3,500, 
and 1,100 trajectories are analyzed using $\Delta t/10^6=$ $30$, $1.8$, 
$1.6$, and $30$; the $\ln [P(t)\Delta t]$ shown are for $t$ values equal 
to $1$, $1/10$, $1/20$, and $1/2$ of that given by the horizontal axis. 
Solid and dashed lines are linear fits for $\epsilon/k_{\rm B}T=$ 
$-1.47$ and $-1.28$.
\\

\noindent 
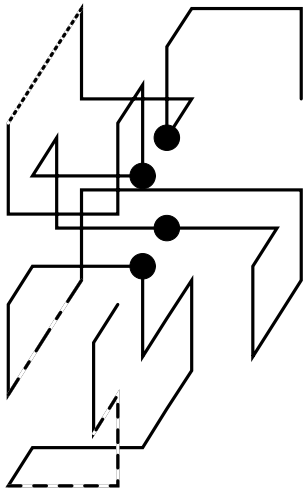
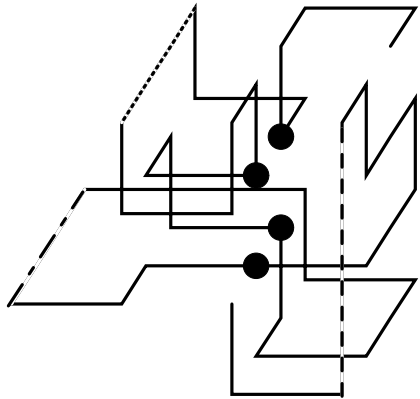
{\bf Fig.~3} $\quad$ 
(a) Ground state of the G\=o model. (b--d) Transiently trapped 
conformations under strongly native conditions (c.f. Fig.~1),
with $Q=39/57$, $41/57$ and $53/57$ respectively. 
The dotted, dashed, dotted-dashed lines and 
filled circles are used to identify monomers (in all four conformations)
that belong to three of the straight edges and the core positions in 
the ground-state conformation (a). The conformation in (d) may
reach the ground state by a simple hinge motion of the dotted-dashed edge. 
However, since such a move is not available in the model [14], the chain now 
must first partially open up before it can access the ground state. 
The trapping effect of (d) is minor compared to that of (b)
and (c) (see Fig.~1).
\\

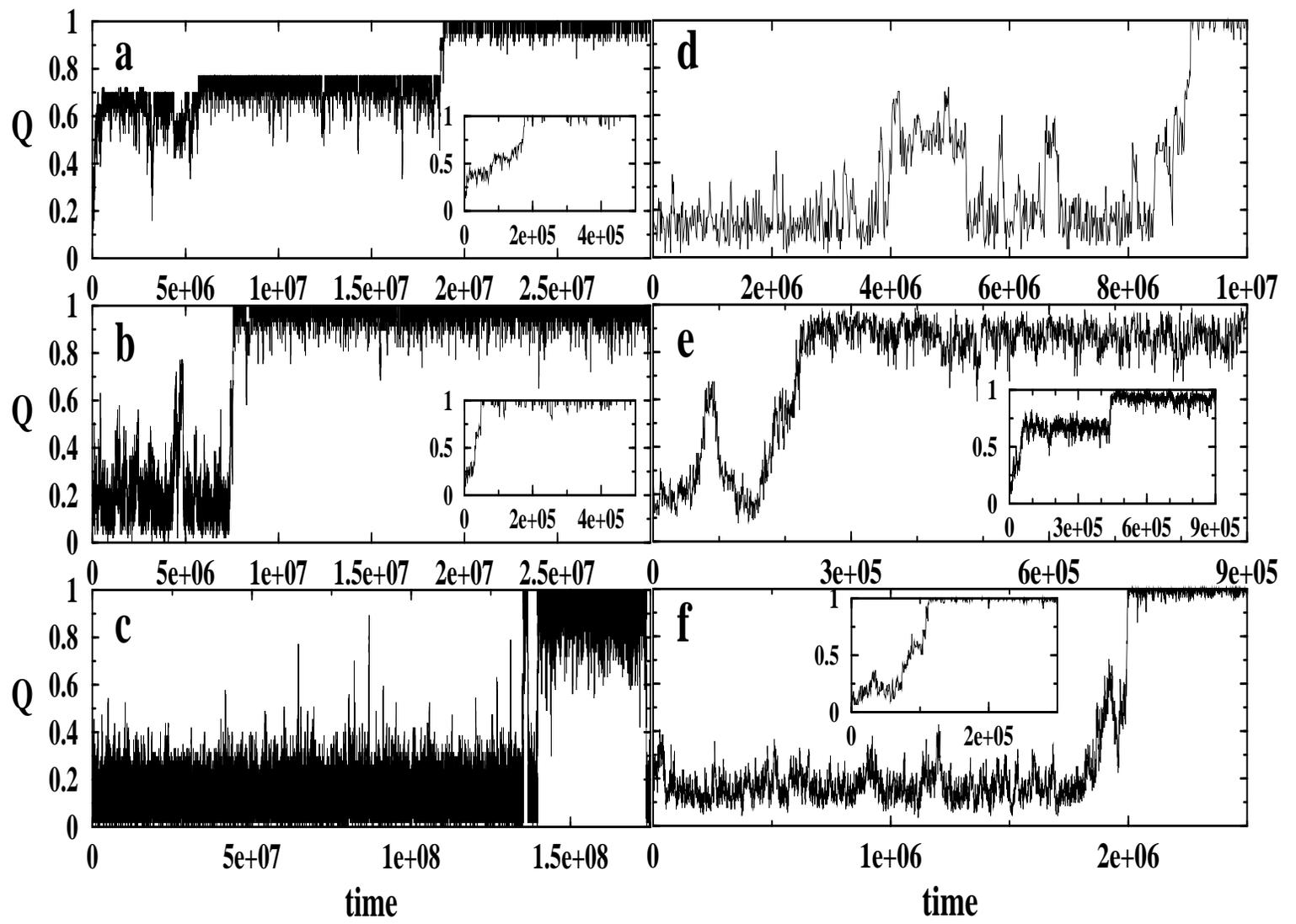
{\bf Fig.~4} $\quad$
(a--c) Folding trajectories of the 48mer G\=o model at 
$\epsilon/k_{\rm B}T=$ $-1.82$ (a), $-1.47$ (b), and $-1.28$ (c). The 
insets in (a, b) each shows a faster folding trajectories at the same 
given $\epsilon/k_{\rm B}T$. (d--f) Typical folding trajectories in
other models for comparison: (d) is from a 55mer lattice model
under mildly native conditions ($\epsilon/k_{\rm B}T=-1.75$) [11].
(e, f) are from the continuum NCS1 without-solvation (e) and 
with-solvation (f) Langevin dynamics models at $T=0.82$ for CI2 [17] 
with $\epsilon=$ $0.88$ (e) and $1.1$ (f). The insets in (e, f) show 
trajectories at the same $T$ but under more strongly native 
conditions at $\epsilon=$ $1.0$ (e) and $1.5$ (f).
\\


\begin{thebibliography}{999}

\bibitem{1}
A. R. Fersht, Curr. Opin. Struct. Biol. {\bf 7}, 3 (1997);          
D. Baker, Nature {\bf 405}, 39 (2000).

\bibitem{2}
S. E. Jackson and A. R. Fersht, Biochemistry {\bf 30}, 10428 (1991).

\bibitem{3}
C. R. Matthews, Methods Enzymol. {\bf 154}, 498 (1987).

\bibitem{4} 
A. Matouschek et al., Nature {\bf 346}, 440 (1990).

\bibitem{5} 
W. A. Houry, D. M. Rothwarf and H. A. Scheraga, Nature Struct. Biol.
{\bf 2}, 495 (1995).

\bibitem{6}
T. Kiefhaber, Proc. Natl. Acad. Sci. USA {\bf 92}, 9029 (1995).

\bibitem{7} 
M. Silow and M. Oliveberg, Proc. Natl. Acad. Sci. USA {\bf 94}, 6084 (1997);
D. E. Otzen et al., Biochemistry {\bf 38}, 6499 (1999).     

\bibitem{8}
R.-A. Chu and Y. Bai, J. Mol. Biol. {\bf 315}, 759 (2002).

\bibitem{9}
R. Li, J. L. Battiste and C. Woodward, Biochemistry {\bf 41}, 2246 (2002).

\bibitem{10}
R. L. Baldwin, Fold. Des. {\bf 1}, R1 (1996).

\bibitem{11}
H. Kaya and H. S. Chan, J. Mol. Biol. {\bf 315}, 899 (2002).

\bibitem{12}
K. W. Plaxco {\it et al.}, Biochemistry {\bf 39}, 11177 (2000).

\bibitem{13}
C. Micheletti {\it et al.}, Phys. Rev. Lett. {\bf 82}, 3372 (1999).

\bibitem{14}
V. S. Pande and D. S. Rokhsar, Proc. Natl. Acad. Sci. USA {\bf 96}, 1273 (1999).

\bibitem{15}
C. Clementi, H. Nymeyer and J. N. Onuchic, J. Mol. Biol. {\bf 298}, 937 (2000).

\bibitem{16}
H. Kaya and H. S. Chan, Proteins Struct. Funct. Genet. {bf 40}, 637 (2000);
Phys. Rev. Lett. {\bf 85}, 4823 (2000).

\bibitem{17}
H. Kaya and H. S. Chan, J. Mol. Biol. {\bf 326}, 911 (2003).

\bibitem{18}
E. Shakhnovich and A. R. Fersht, Curr. Opin. Struct. Biol. {\bf 8}, 65 (1998);
V. S. Pande {\it et al.}, Curr. Opin. Struct. Biol. {\bf 8}, 68 (1998);
R. L. Baldwin and G. D. Rose, Trends Biochem. Sci. {\bf 24}, 77 (1999);
S. W. Englander, Annu. Rev. Biophys. Biomol. Struct. {\bf 29}, 213 (2000).

\bibitem{19}
J. D. Bryngelson {\it et al.}, 
Proteins Struct. Funct. Genet. {\bf 21}, 167 (1995);
K. A. Dill {\it et al.}, Protein Sci. {\bf 4}, 561 (1995); D. Thirumalai and
S. A. Woodson, Acc. Chem. Res. {\bf 29}, 433 (1996).

\bibitem{20}
H. S. Chan and K. A. Dill, Proteins: Struct. Funct. Genet. {\bf 30}, 2 (1998).

\bibitem{21}
M. S. Li and M. Cieplak, Eur. Phys. J. B {\bf 14}, 787 (2000).

\bibitem{22}
J. N. Onuchic {\it et al.}, Fold. Des. {\bf 1}, 441 (1996).

\bibitem{23}
N. D. Socci, J. N. Onuchic and P. G. Wolynes, J. Chem. Phys.
{\bf 104}, 5860 (1996).

\bibitem{24}
J. Sabelko, J. Ervin and M. Gruebele, Proc. Natl. Acad. Sci. USA
{\bf 96}, 6031 (1999).

\bibitem{25}
V. I. Abkevich, A. M. Gutin and E. I. Shakhnovich, J. Chem. Phys.
{\bf 101}, 6052 (1994).

\bibitem{26}
M. S. Cheung, A. E. Garc{\'\i}a and J. N. Onuchic, Proc. Natl. Acad. Sci.
USA {\bf 99}, 685 (2002).

\bibitem{27}
R. Miller {\it et al}., J. Chem. Phys. {\bf 96}, 768 (1992).

\end{thebibliography}
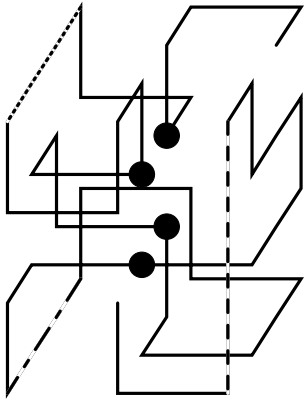
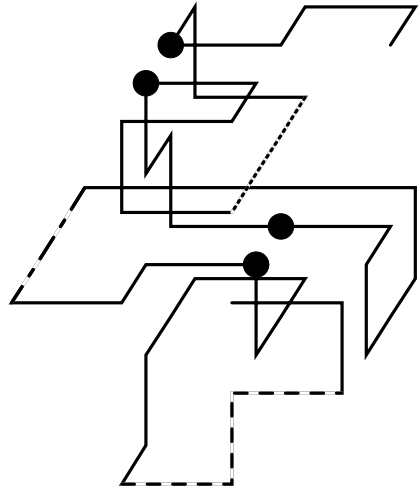
\end{document}